# 3-Dimensional residual neural architecture search for ultrasonic defect detection

Shaun McKnight[1*], Christopher MacKinnon[1], S. Gareth Pierce[1], Ehsan Mohseni[1], Vedran Tunukovic[1], Charles N. MacLeod[1], Randika K. W. Vithanage[1], Tom O'Hare[2]

[1]*Sensor Enabled Automation, Robotics, and Control Hub (SEARCH), Centre for Ultrasonic Engineering (CUE), Electronic and Electrical Engineering Department, University of Strathclyde, Glasgow, UK*

[2]*Spirit AeroSystems, Belfast, UK*

## Abstract

This study presents a deep learning methodology using 3-dimensional (3D) convolutional neural networks to detect defects in carbon fiber reinforced polymer composites through volumetric ultrasonic testing data. Acquiring large amounts of ultrasonic training data experimentally is expensive and time-consuming. To address this issue, a synthetic data generation method was extended to incorporate volumetric data. By preserving the complete volumetric data, complex preprocessing is reduced, and the model can utilize spatial and temporal information that is lost during imaging. This enables the model to utilise important features that might be overlooked otherwise.

The performance of three architectures were compared. The first two architectures were hand-designed to address the high aspect ratios between the spatial and temporal dimensions. The first architecture reduced dimensionality in the time domain and used cubed kernels for feature extraction. The second architecture used cuboidal kernels to account for the large aspect ratios. The evaluation included comparing the use of max pooling and convolutional layers for dimensionality reduction, with the fully convolutional layers consistently outperforming the models using max pooling. The third architecture was generated through neural architecture search from a modified 3D Residual Neural Network (ResNet) search space. Additionally, domain-specific augmentation methods were incorporated during training, resulting in significant improvements in model performance for all architectures. The mean accuracy improvements ranged from 8.2% to 22.4%. The best performing models achieved mean accuracies of 91.8%, 92.2%, and 100% for the reduction, constant, and discovered architectures, respectively. Whilst maintaining a model size smaller than most 2-dimensional (2D) ResNets.

## 1. Introduction

Composites are versatile materials that are widely used in many industries due to their superior mechanical properties such as corrosion resistance, specific strength, and specific stiffness. Carbon Fibre Reinforced Polymer (CFRP) is a widely used composite in the aerospace industry making up over 50 wt%  for the two most recent long-range aircraft, the Airbus A350 and the Boeing 787, and up to 70-80 wt% for private jets and helicopters [1]. CFRP is fabricated by layering multiple carbon ply sheets and curing them with a thermoset polymer. The anisotropic nature of these composites can be engineered to meet specific structural requirements, making them ideal for high-performance applications [2]–[10]. However, the manufacturing process can introduce defects in the composite, compromising its integrity and performance [2], [3], [5], [8], [9], [11], [12]. Defects can range from delamination and cracks to foreign object inclusions, fibre distortions, and porosity [7], [12]. As the use of composites in safety-critical parts continues to rise, the detection, characterization, and quantification of defects become increasingly important [5].

Non-Destructive Evaluation (NDE) refers to a suite of techniques employed to inspect components without causing any damage. Radiography, thermography, electromagnetic methods, and ultrasound are among the most widely used NDE techniques. These methods allow inspection of components with varying levels of complexity and size. The choice of NDE technique depends on the nature of the component and the defects to be detected. The application of these NDE techniques has significantly improved the reliability and safety of various structures and components across numerous industries.

Ultrasonic Testing (UT) is a versatile technique that can be used to inspect components made of various materials, including composites, and is based on the transmission, propagation, and reception of ultrasonic waves. UT has become widely adopted and standardized for volumetric inspection in the aerospace industry due to its relatively easy and hazard free implementation compared to radiography and its ability to detect a wide range of volumetric defects [3], [7], [10], [11]. In UT, sound waves are excited on the surface of a component, and the reflected/scattered wave from internal scatterers can provide valuable information about the volumetric discontinuities of the component. Currently, phased arrays are the preferred technology for generating the initial sound wave owing to their operational flexibility. Phased arrays employ independently controllable UT elements that enable more complex electronic scanning and imaging, such as beam steering, dynamic depth focusing, and variable sub-apertures [9]. By controlling each individual element (or sub-aperture of elements) of a linear phased array, depth-wise sectional images (B-scans) can be created in a single scan (Figure 1, Figure 2 (a)Figure 2). When combined with mechanized scanning perpendicular to the length of a linear phased array, complete 3-dimensional (3D) volumetric scan data of components can be generated by stacking multiple individual B-scans together at known positions (Figure 2 (b)). This technique is highly valuable for assessing the structural integrity of large and complex components and has significant implications for the reliability and safety of aerospace structures.

UT data is commonly visualized as images, either by selecting a B-scan directly, or as an amplitude or time of flight C-scan; where either the maximum response amplitude or the time index of the maximum response amplitude within the volume is imaged to produce a top-down section view across the sample [13].

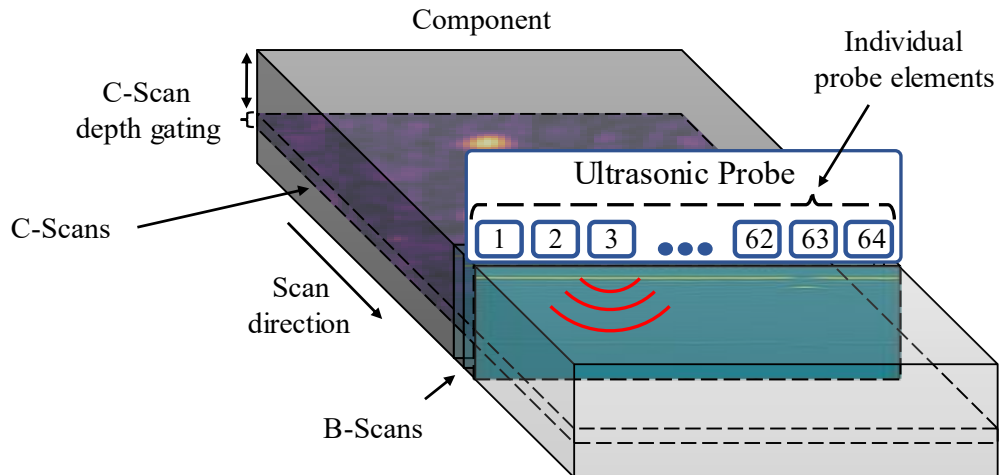

*Figure 1: Demonstration of how individual probe elements can make up a linear phased array which can produce B-scan and C-scan images.*

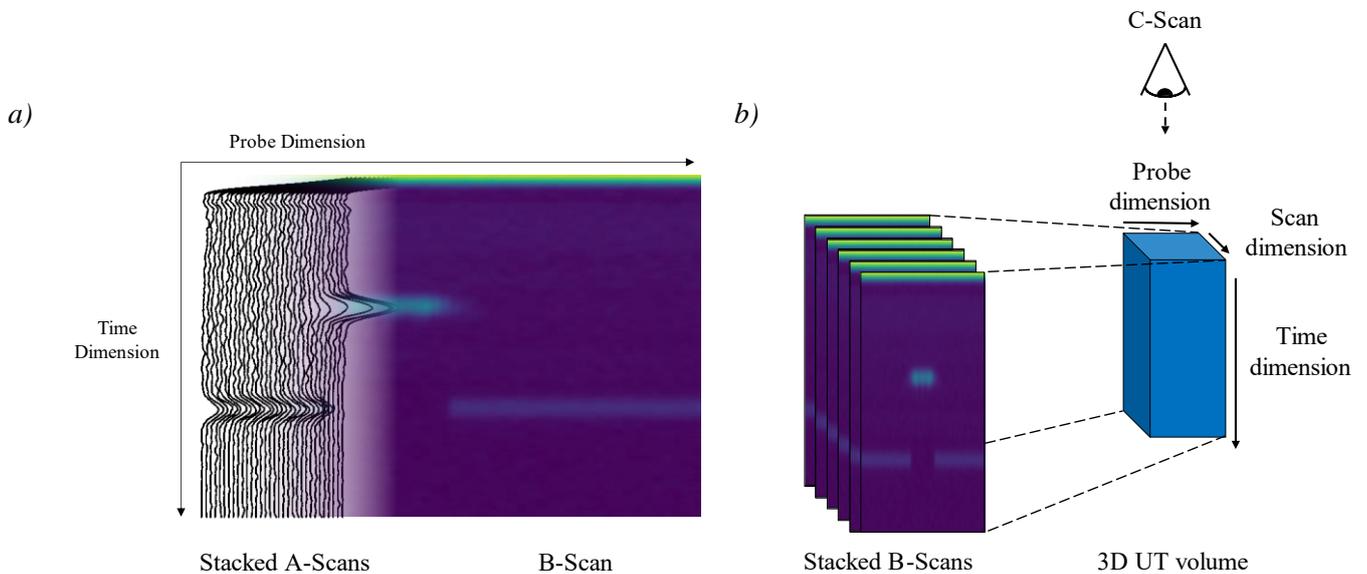

*Figure 2: a) Representation of how A-scans are stacked to form B-scans. b) How B-scans are stacked to create a full UT volume.*

The integration of robotics into NDE has revolutionized large-scale inspection processes by enabling efficient automated inspection of large components [14]. While robotic scanning offers greater flexibility and significantly reduces scan time compared to manual scanning, the interpretation of results remains a tedious and time-consuming task in industry. To interpret results in line with existing standards, there is a requirement for highly trained and qualified operators [10], [15]–[19]. Despite the significant improvements brought about by robotic NDE, the need for expert human interpretation of results persists. This highlights the need for further research and development of automated data interpretation techniques that can supplement or even replace human interpretation, to improve the efficiency and reliability of NDE in various industries. By reducing the dependence on human interpretation, automation can potentially enhance the consistency, repeatability, and traceability of the NDE processes, while reducing inspection time and costs.

The interpretation of UT scan results by human operators presents two significant drawbacks, namely, poor time efficiency and the risk of human error [17]. Low levels of automation for data interpretation is feasible for mass-produced parts with precisely known geometries, but this approach typically relies on hard-coded features such as predefined time-gating, filtering, and amplitude thresholding, which may not be adequate for complex tasks such as changes in manufacturing conditions, variations in geometry, or defect characterisation [19]. Hence, the development of a Deep Learning (DL) approach that can automate the interpretation of complex UT results and work seamlessly with robotic inspection to substantially enhance the quality and efficiency of large component inspection. This would lead to shorter signal interpretation time and faster UT automation uptake in aerospace and other industries, where DL has been identified as a key requirement for transitioning from low to high levels of industrial automation [19].

Despite the potential benefits of applying DL techniques to ultrasonic signal analysis for composite components, its uptake has been limited [19]. Shortage of training data is one of the main challenges that hinders research developments in this area. This shortage, combined with industrial concerns about the interpretability and compliance with standards of DL models, has presented challenges for the effective use of DL techniques. As a result, the adoption of DL in UT signal analysis for composite components has been slow, despite its promising potential to enhance the accuracy and efficiency of defect detection and characterization.

Synthetic datasets are widely used in Machine Learning (ML) to augment small training datasets [20] and they have been successfully implemented for UT of composites with encouraging results for 2-dimensional (2D) classification of C-scan images [21]. Part of this work builds upon the work in the previous paper to extend one of the synthetic data generation methods to make it applicable for full 3D volumes. The synthetic datasets are based on simulations from semi-analytical physics based software that has been shown to produce experimentally accurate defect responses [22], [23]. This software offers a less computationally expensive alternative to Finite Element Analysis (FEA), allowing for the simulation of composite responses based on bulk material properties [24].

When ML is used to interpret UT NDE data in literature, it is typically applied to interpret A-scan time traces or 2D images constructed from A-scans [25]–[31]. Compared to B-scans, A-scans lack all spatial information and nowadays, they are rarely used alone to characterize defects by human operators since the introduction of phased arrays. C-scans preserve detailed spatial information, however constructing the 2D image from the volumetric data necessitates the compression of temporal information. Whilst C-scans excel in capturing intricate spatial details, their need for temporal compression results in minimal representation of through-depth features. Compression of A-Scans to C-Scans often removes useful features such as the backwall response, which can be important when detecting defects with a low reflective index such as porosity [32]. Furthermore, to produce C-scan images appropriate gating must be applied to remove the front wall surface response. This can be challenging when trying to detect near-surface defects. In the aerospace industry, operators typically start with a C-scan to gain a complete picture of defect responses and then move to analysis of B-scans for further information about the nature of the responses [33], [34]. Whilst current ML approaches in literature make use of data in formats that are easily interpreted by humans (images or time-traces), ML algorithms are not limited to image-level analysis and have proved very capable at interpreting 3D volumetric data [35], [36]. By implementing algorithms capable of volumetric interpretation, we retain all spatial and depth information, this gives the algorithms more relevant features to learn from and removes the need for image pre-processing and gating.

Convolutional Neural Networks (CNNs) have been used effectively for decades in a wide variety of image and volumetric analysis tasks with models such as ResNet typically having tens of millions of parameters [37], and are still widely used as backbones or standalone architectures [38]. However, these networks are typically applied to data of similar dimensions, or data which has been scaled to give even dimensionality of each axis. UT data has extreme aspect ratios due to the difference in requirements of sample rate in the spatial and time dimensions. Compressing the data in the time dimension to match the spatial dimension, normally dictated by the sub-aperture pitch and the scan acquisition rate, would result in a substantial loss of depth information. Alternatively, the spatial dimensions could be upscaled to match the number of samples in the time dimension, but this is highly inefficient, creating data instances that would require large amounts of memory, and would make training intractable. Therefore, retaining the original dimensionality and aspect ratio of the UT data is highly preferable. Using CNNs to interpret images with high dimensionality is not new and the use of rectangular kernels instead of square kernels in CNNs has given positive results for classification of speech signals, which have high aspect ratios when represented as spectrogram images [39]. This paper makes use of a similar approach for volumetric data.

Network architecture design is a key component of effectively leveraging machine learning techniques. Traditionally, network design heuristics and 'rules of thumb' would be used, in tandem with domain expert knowledge to construct a specific architecture. Automatic architecture design or Neural Architecture Search (NAS) is a development on this approach where a practitioner can leverage compute to aid the process of architecture selection. This process, which can be considered a subset of hyperparameter optimization, generally involves an iterative process of selecting, training, and evaluating architectures. In its simplest form, a 'Random Search' involves repeating the above process until some threshold or limit in terms of performance or computation budget is reached. More complex approaches to NAS often focus on efficient model evaluations, making use of proxy evaluation methods [40], [41] or efficient sampling algorithms [42], [43] attempting to make the largest improvement with each evaluation.

This paper presents a comparative analysis of the performance achieved by two hand-designed architectures and a searched architecture for defect detection in volumetric ultrasonic data. Additionally, notable contributions of this study to knowledge in the field encompass the introduction of domain-specific augmentations, which exert a substantial impact on the classification performance. Furthermore, synthetic data generation techniques are employed to generate 3D UT datasets from semi-analytical simulations, effectively addressing one of the prominent challenges encountered in the application of deep learning for NDE: the scarcity of effective training data.

### 1.1. Pipeline overview

In this work, the automated data interpretation is simplified by inspecting the complete volumetric data, eliminating image processing steps like gating to remove front and back wall responses, while preserving all spatial and temporal information. Whilst the models are trained on synthetic data, they are tested using experimentally collected UT data from samples with manufactured defects that aim to mimic delamination's. Manufactured defects are commonly used in literature to act as test cases and qualify NDE techniques and operators where naturally occurring defects are not always available [6], [27], [28]. An overview of the simulation and deep learning pipeline is presented in Figure 3. Figure 3 also shows how NAS can fit into this process, with Figure 18 providing a more detailed overview of the NAS pipeline.

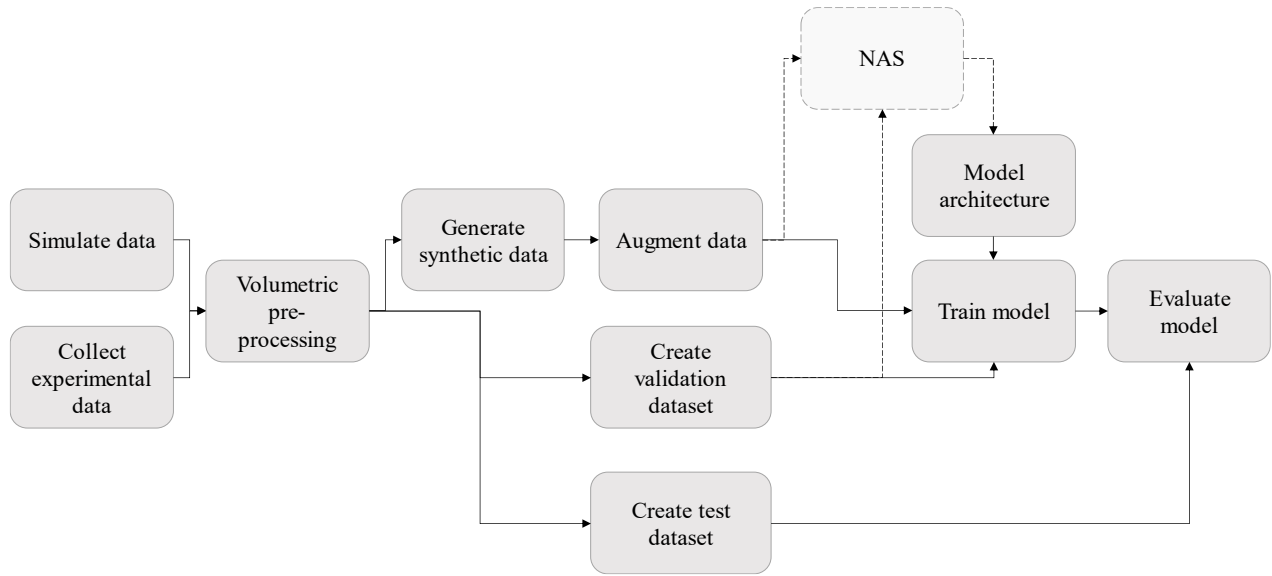

*Figure 3: Overview of the pipeline for automated volumetric UT classification.*

## 2. Acquisition of experimental test data and synthetic training data

### 2.1. Experimental data collection

Experimental ultrasonic data was acquired from CFRP samples, both with and without artificially introduced defects, to serve as test data. To imitate delamination defects, which are one of the most common defects in composites [31] and a significant life-limiting failure mode [44], flat-bottom holes were drilled from the backside of the samples. Prior to introducing defects, clean scans of each sample were taken to form a defect-free test set. The use of the same CFRP base sample ensured that the trained models learned defect-specific responses rather than the underlying properties of different composite samples.

Composite samples measuring 254.0 x 254.0 x 8.6 mm (W x D x H) were provided by Spirit AeroSystems and were manufactured to the BAPS 260 specification with woven fabric, and Cycom 890 resin using a vacuum assisted resin transfer molding process. Of the three samples, two contained defects. The first sample contained 15 flat-bottom holes measuring 3.0, 6.0, and 9.0 mm in diameter, with each defect drilled to depths of 1.5, 3.0, 4.5, 6.0, and 7.5 mm from the front surface. The different defect sizes were spaced 30 mm apart with different depth defects spaced 35 mm apart. The second sample contained 25 flat-bottom holes, drilled to the same depths as the first sample but with additional defect diameters of 4.0 and 7.0 mm, as shown in Figure 4. All defects were manufactured to tolerances in depth of +/- 0.3 mm and in diameter of +/- 0.2 mm.

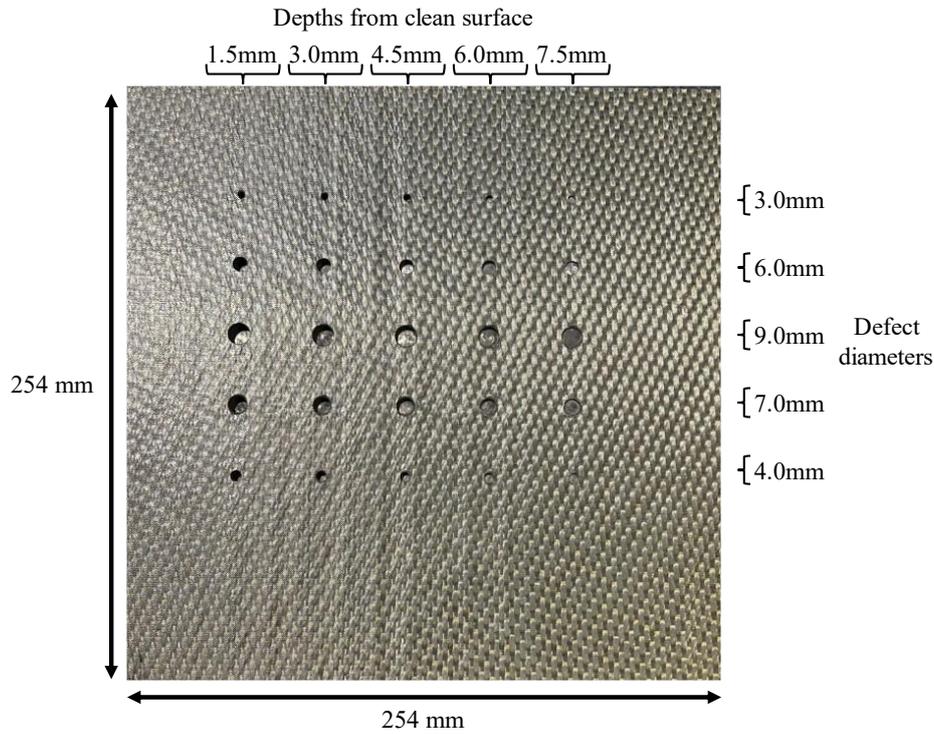

*Figure 4: The composite test sample showing 25 Flat-Bottom Holes.*

The ultrasonic data was acquired using a robotically deployed unfocused linear phased array. The array used was an Olympus Inspection Solutions RollerFORM-5L64 [45], which had a central frequency of 5 MHz and was made up of 64-elements with a pitch of 0.8 mm and elevation of 6.4 mm. The elements were driven at 100 V with a receiver gain of 22.5 dB. The pulse repetition frequency was set to collect a B-scan every 0.8 mm with a scan speed of 10 mm/s which was controlled using a fully automated robotic system built around a KUKA KR 90 R3100 extra HA industrial robot (Figure *5*) Figure 4Figure *5*[46]. Robotic scanning enabled the concatenation of encoded B-scans to form volumetric datasets. To ensure a steady coupling of the roller-probe to the surface of the component and consistent transfer of acoustic wave energy into the sample at different scanning positions, Force-Torque compensation was used to control the distance from the samples surface with feedback from the force axis perpendicular to the sample. This was accomplished with integration of a Schunk GmbH & Co. FTN-GAMMA-IP65 SI-130-10 Force-Torque sensor, mounted between the robot's flange and the roller-probe. This ensured a constant scanning force of 70 N which maintained consistent tyre compression throughout the scan. Water was used as an acoustic couplant in the scanning process. This data acquisition setup is widely used in industry and has been employed for data collection on large composite aerospace components [47].

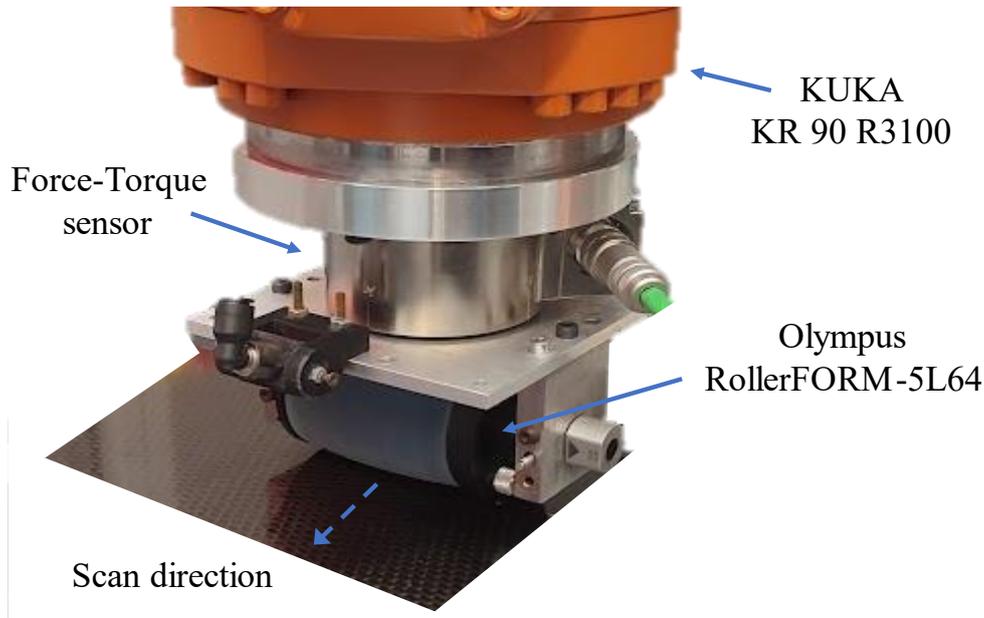

*Figure 5: Overview of the experimental scan setup of KUKA KR90, Force-Torque sensor, and ultrasonic roller probe used for data acquisition.*

### 2.2. CIVA Simulations

Due to the lack of available experimental training data, a simulated dataset was constructed for training. This was done using CIVA, a semi-analytical physics-based commercial NDE simulation software [48]. CIVA has the ability to accurately model wave propagation and its interaction with defects, which has been experimentally validated for UT [22], [23]. Additionally, the software is computationally efficient when compared to other alternatives such as Finite Element Analysis (FEA). The full control of the simulated domain enabled the modelling of similar defects and material properties to the experimental domain. However, the use of semi-analytical software instead of FEA had limitations in that the software was unable to model responses from ply interactions and lacks noise seen in experimental data. As a result, differences existed between the simulations and measured experimental responses, leading to the use of the synthetic data generation steps discussed in Section 2.4 to reduce the differences between simulated and experimental domains. For further information on the difference between the simulated and experimental domain and the need for accurate synthetic data, please refer to the previous work on this topic [21].

To setup the simulation, the individual layers of composite were constructed and used to generate equivalent homogeneous material properties of the experimental CFRP samples. A single ply layer was constructed and alternated repeatedly with 8 layers at orientations of 0, 45, -45, and 90 degrees to match the experimental sample as closely as possible. The resulting multilayer structure was homogenized to be consistent with a homogeneous medium having mechanical properties equivalent to those of the multi-ply composite, with the fiber density set to 50% best match the experimental sample's density of 1440 kg/m3.

For running multiple, sequential simulations, a parametric study was set up, using the composite bulk properties previously calculated and varying the diameter and depth of defects. Flat bottom hole defects were simulated with diameters from 3.0 mm to 15.0 mm, increasing every 0.5 mm, with varying depths from 1.5 mm to 7.0 mm from the surface, in increments of 1.5 mm. A defect-free simulation was also run to provide the basis for defect-free synthetic data. Both the front and back wall surface reflections were included in the model. The full simulations took less than 15 hours on a desktop computer with a 24-Core 3.79 GHz CPU and 128 GB of memory.

### 2.3. Signal processing and dataset generation

The resolution of the UT data in the array dimensions was constrained by the element pitch, and the scan width was restricted by the number of elements in the array. This limited the inspection data to 64 voxels in the array dimension. To match this, 64 B-scans were selected in the scan dimension to create cuboidal datasets (Figure

2(b)). The distance between the elements was 0.8 mm, and the robotic scanning speed was regulated with the pulse repetition frequency to ensure a B-scan offset of 0.8 mm. This enabled the generation of volumes with square voxels in the spatial domains, along both the probe and scan directions. By utilizing this approach, the study was able to achieve a standardized volumetric resolution that was consistent throughout the dataset.

The data collected in both experimental and simulated domains was in the form of radio frequency A-scans, also known as amplitude scans. In order to obtain 3D volumetric datasets from these sources, a number of signal processing steps were performed. Initially, the A-scans were centered at zero amplitude and enveloped by taking the absolute of the Hilbert transform, as shown in Figure 6 (a). The Hilbert transform was used to obtain the analytical signal, which is useful for calculating the instantaneous response of a time series. This approach is a standard signal processing technique used when generating C-scan images from time series ultrasonic data. Subsequently, each volumetric dataset was normalized between 0 and 1 by dividing by its maximum peak amplitude.

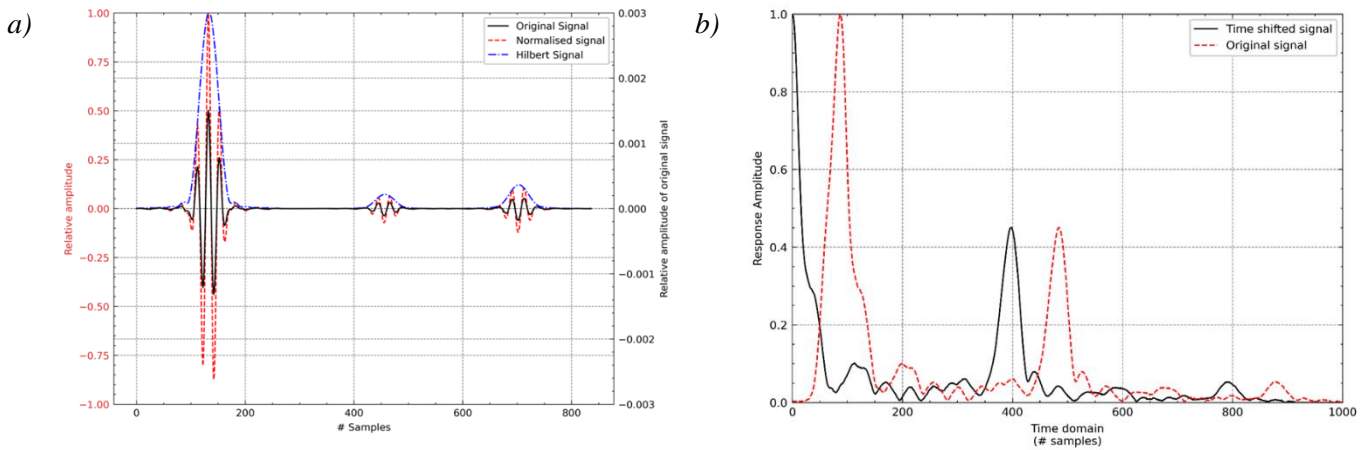

*Figure 6: a) Example of relative amplitude response from simulations, normalized signal, and Hilbert transform, applied to the original signal. b) Demonstration of how individual A-scans are time shifted to the front wall response.*

Once the data was normalized, the offset in the time domain was compensated for by aligning the peak front wall response to the origin. This made sure that features were correctly aligned in the time domain and helped to account for any variability in the acoustic path length between individual transducers and the surface of the sample. Figure 6 (b) Shows how the time shifting was done for an individual A-scan with the Hilbert transform applied. Figure 7 shows the effect of this on the complete ultrasonic volume.

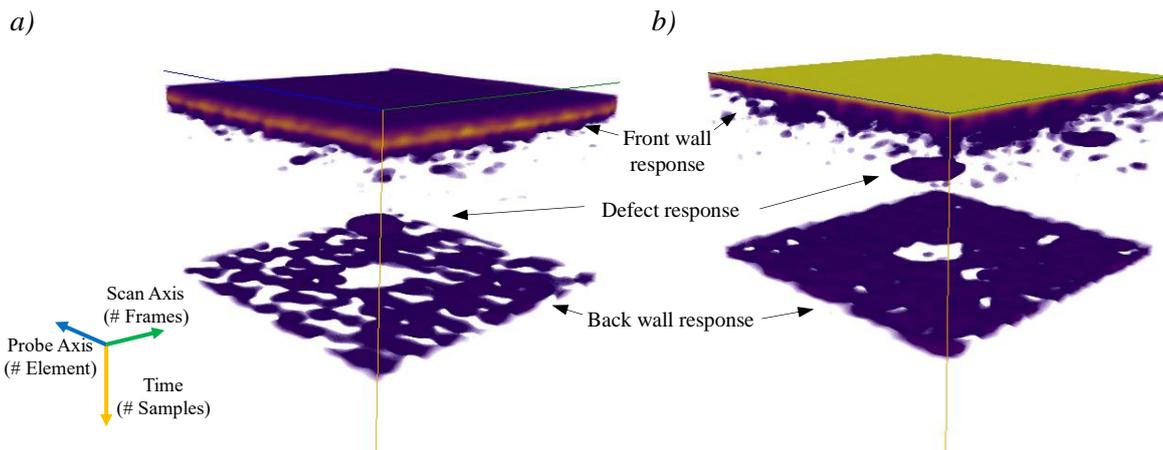

*Figure 7: a) Volumetric data with Hilbert transform applied only. b) Volumetric data with time shifting to the central response of the front wall peak. Both figures have been thresholded to remove the lowest 10% of amplitudes to aid in visual clarity.*

### 2.4. Synthetic data generation method

Our previous studies have shown that semi-analytical simulated data alone is not representative enough of the experimental domain [21]. Therefore, there is a need for methods of translating the simulated domain closer to the experimental domain. Fully statistical methods of generating noise are advantageous as they can be re-sampled continuously to keep generating unique noise profiles which are in line with experimental data. In this paper, we extend previous work in generating 2D synthetic images, and propose a new approach for adding noise to complete volumetric UT data.

The previous study concluded [21] that A-scan level noise was the best fully generative statistical method for adding noise. Additionally, all the other approaches, except for the simulated A-scan noise, introduced noise at an image level, which is intractable for volumetric data. To adapt the methodology described in the original paper for the analysis of full volumetric data, unique noise profiles for each A-scan were generated and subsequently summed with the simulated responses past the front wall. To temporally align the responses, the time shift of the front wall response was performed when generating the noise distributions from experimental data, and when combining the simulated responses with the generated noise profiles (Figure 6 (b)).

Figure 8 shows an example of the addition of noise on simulated data at an A-scan level and Figure 9 demonstrates this for a complete ultrasonic volume. The statistical noise distributions of the A-scans were calculated from a separate hold out sample with the same layup and thickness as the test samples. For further details on building up the noise profiles, please refer to the previous work [21].

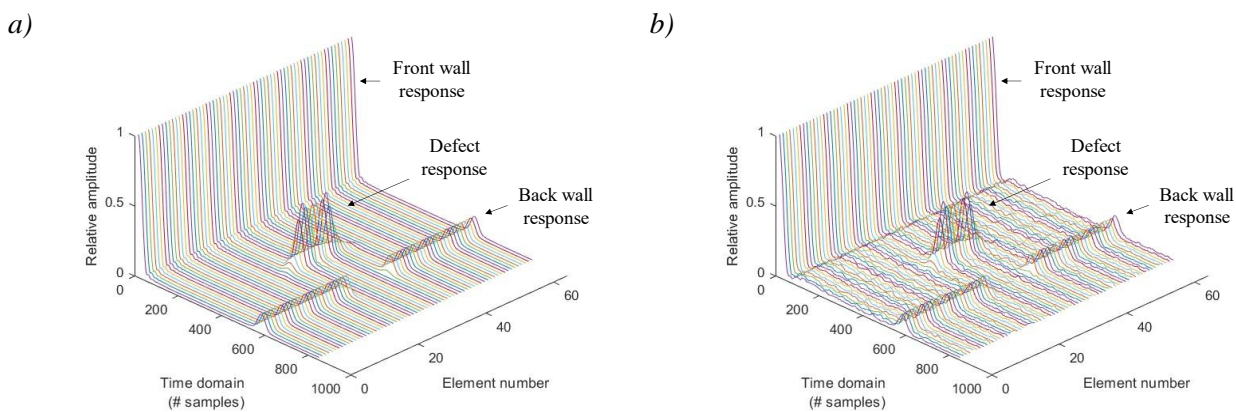

*Figure 8: a) A frame of 64 simulated A-scans for a simulated defect response. b) the corresponding A-scans with synthetically added noise for the same defect response.*

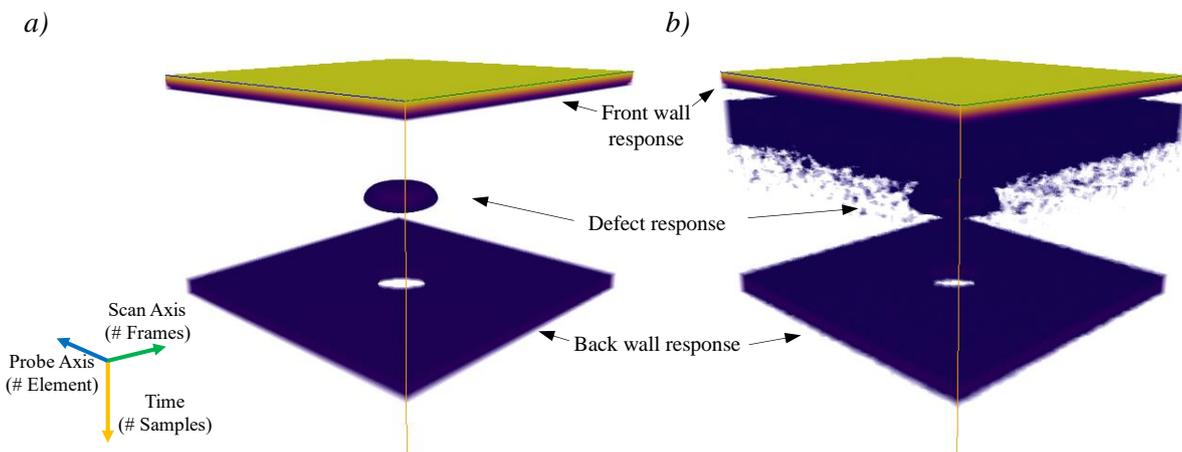

*Figure 9: a) Complete ultrasonic volume of simulated A-scans for a defect response. b) the corresponding synthetically noised volume for the same defective response. Both figures have been thresholded to remove the lowest 10% of amplitudes to aid in visual clarity.*

A summary of the datasets generated from the experimental and synthetic data is given in Table 1.

*Table 1: Summary of the datasets produced.*

| Data source | Dataset | Number of datapoints |
| --- | --- | --- |
| Simulated defect responses (300 Flat-Bottom Holes) | Synthetic defective train | 300 |
| Simulated defect free response | Synthetic defect free train | 300 |
| Experimental defect reference sample (15+25 Flat-Bottom Holes) | Defect test (70%) | 25 |
|  | Defect validation (30%) | 15 |
| Experimental defect free reference sample | Defect free test (70%) | 25 |
|  | Defect free validation (30%) | 15 |

### 2.5. Augmentation methods for synthetic training data

The generalizability of ML models is a critical aspect of their performance. One approach to improve generalizability is to augment the training data. Augmenting the training data makes the task more challenging by adding noise at the training stage, reducing the likelihood of overfitting, and often improves performance in the target domain. This is particularly important when the target (experimental) domain is different from the training (synthetic) domain.

In this study, we introduce two types of augmentation that were generated online for each minibatch during training. These augmentations aim to mimic the inter-element response variability observed within the UT probes used for data collection.

The first type of augmentation is related to the magnitude of response measured by the UT elements, which varies due to many factors not included in the simulation, such as manufacturing tolerances of the sample and the UT array probe, wear and tear of the probe and electrical wires/connections, or inter-layer multiple scattering of the sound waves. To mimic these whilst preserving the correct normalization, each A-scan was scaled by a constant past the front wall. The scale factor was sampled from a uniform distribution to give a scale factor between 80-120%. An example of this is given in Figure 10 (a).

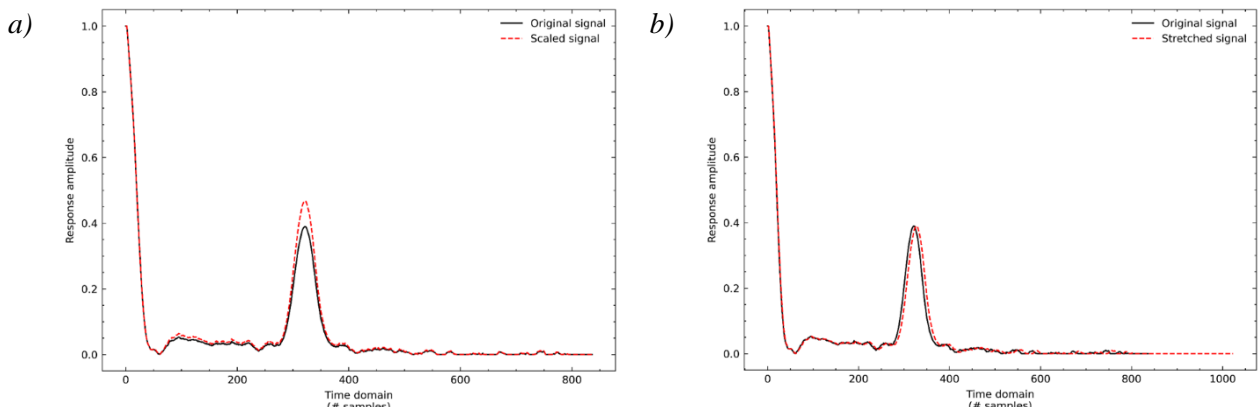

*Figure 10: a) Example of how scaling augmentation is done on an individual A-scan. b) Example of how dilation augmentation and padding is completed for an individual A-scan.*

The second type of augmentation mimics any changes in ultrasonic travel time seen by different elements. This can be caused by a variety of factors, such as variations in component sound speed due to the anisotropic nature of composites, departure from central frequency for certain elements, etc. To simulate this 1-D interpolation was used to randomly stretch or compress the signal in the time domain. The dilation amount was randomly sampled from a uniform distribution for each A-scan up to $\pm 15$ samples. For compressed signals, any trailing empty space was padded with zeros. An example of this is given in Figure 10 (b).

By introducing these augmentation methods, we aim to improve the generalizability of the models to the experimental domain. The online nature of these augmentations means that they can be easily incorporated into the training process without the need for additional data collection or pre-processing steps. To ensure consistent length of data in the time domain, each A-scan was padded with zeroes to a length of 1024 samples during training.

## 3. Network Architectures

In this paper we investigated the performance of three different 3D CNN architectures for binary classification of 3D defect and defect free UT data with extreme aspect ratios. For low aspect ratios CNNs typically make use of square or cuboidal kernels which are appropriate for their equal (or near equal) aspect ratios. The use of CNNs on data with more extreme aspect ratios is less common and is particularly extreme for UT data between the time and the spatial domains, with an aspect ratio of 16. To address this, we designed two custom architectures which aim to tackle the extreme aspect ratio problem in different ways. The first approach aimed to separate the aspect ratio reduction and feature extraction, whilst the second aimed to combine the aspect ratio reduction and feature extraction. For each custom architecture we experimented with the use of strided convolutions or max pooling for dimensionality reduction. For direct comparison the architectures were kept as similar as possible, with the same number of layers, channels, and activation function (Leaky ReLU). We also utilized neural architecture search to develop a third architecture for comparison. For each model Adam optimizer [49] was used with a constant learning rate of 0.001, $\beta_1$ of 0.9 and $\beta_2$ of 0.999. A batch size of 8 and binary cross-entropy loss was used with sigmoid activation on the final layer for classification.

Due to the small amounts of experimental test data, there was a likelihood of noisy results during both training and testing phases. To mitigate this, each model was retrained ten times with varying random initializations, and their individual results were averaged across the performance metrics. This gives a better representation of the model's performance by averaging out any noisy results due to the small datasets. During the training phase, a randomly selected validation set of experimental data was used to monitor the model's performance and prevent overfitting. The models' parameters were recorded after every ten batches, and the validation set was evaluated using binary cross-entropy loss. The model parameters with the lowest validation loss were used to evaluate the classification performance on the test set. This approach ensured that the final model's defect detection performance was evaluated using the parameters that had the best ability to generalize to the target domain, as opposed to the model that had overfit to the synthetic domain.

### 3.1. Evaluation metrics

To quantitatively assess the binary classification performance of each network, average mean accuracy, precision, recall and F1 scores were calculated according to Equations 1-4.

$$Accuracy = (TP + TN) / (TP + TN + FP + FN) \qquad 1$$

$$Precision = TP / (TP + FP) \qquad 2$$

$$Recall = TP / (TP + FN) \qquad 3$$

$$F1 = (2 \times Precision \times Recall) / (Precision + Recall) \qquad 4$$

Where TP is true positive, TN is true negative, FP is false positive, and FN is false negative, with positives being the presence of a defect. Each result was individually averaged using a simple mean across the 10 training cycles.

### 3.2. Architecture 1 "Reduction": Separation of reduction and feature extraction

The first architecture (Figure 11), herein referred to as reduction, reduced the dimensionality in the time domain to match the spatial dimensions of the data with the use of two reduction blocks with either strided convolutions or max pooling layers (Figure 12). Subsequently, three feature blocks made up of equal-sized cuboidal kernels and max pooling or strided convolutional layers were applied to extract features from the data (Figure 13), as is done with traditional CNNs. Global average pooling was used to reduce the number of features, for classification. Batch normalization and dropout were used for regularization, as given by the architecture diagram in Figure 11. When dimensionality reduction was performed using convolutional layers, the estimated total size of the network was 1.54 M parameters. On the other hand, if max pooling was used the estimated total size of the network was 0.60 M parameters.

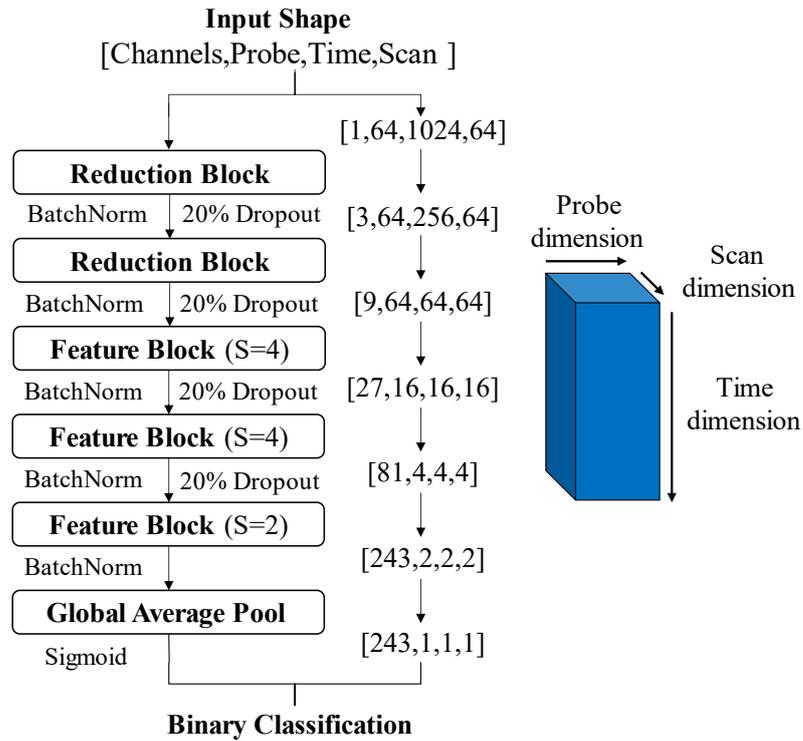

*Figure 11: Network architecture for the reduction model.*

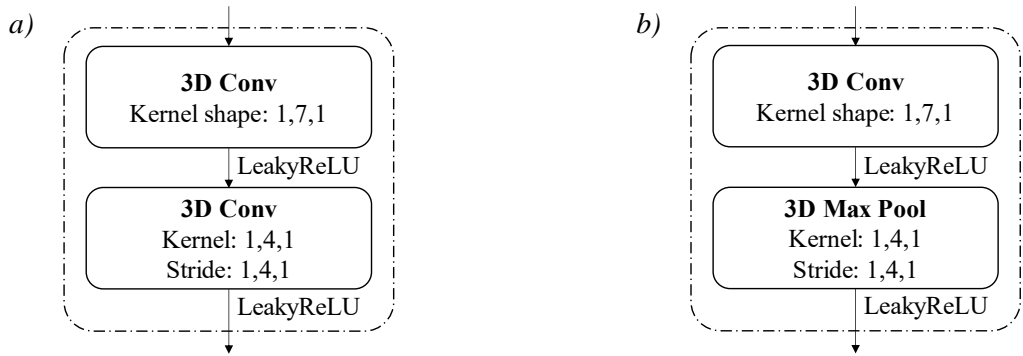

*Figure 12: Reduction blocks with either convolutional (a) or max pooling (b) layers for dimensionality reduction.*

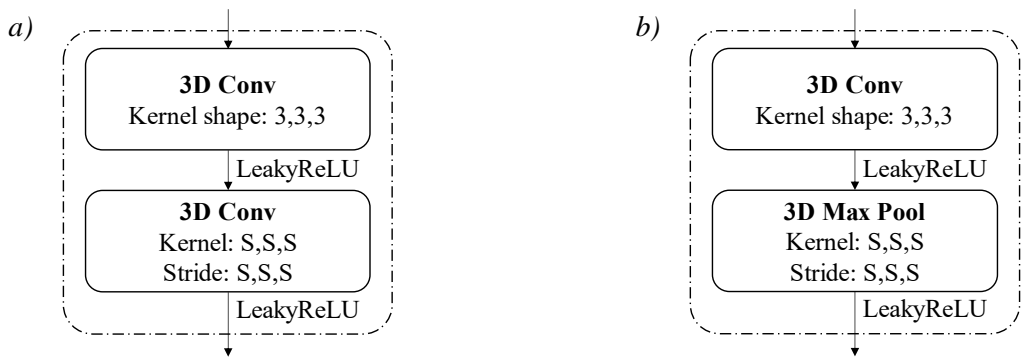

*Figure 13: Feature extraction blocks with either convolutional (a) or max pooling (b) layers for dimensionality reduction.*

### 3.2.1. Classification results

Table 2 provides a summary of classification results for the reduction architecture. The use of fully convolutional layers for dimensionality reduction significantly improved classification performance, with much better performance without data augmentation during training than using max pooling with augmentation. The increased classification performance did come at a computational cost, with the model which used max pooling 39% smaller in parameter size compared to the fully convolutional model. The use of data augmentation during training resulted in a significant increase in classification performance for both dimensionality reduction methods. With a mean accuracy increase of 9.8% for the fully convolutional model and 31.8% for the max pooling model. Despite converging to a loss minimum during training on synthetic data, when using max pooling without data augmentation the model failed to classify experimental volumes almost entirely and demonstrated a level of accuracy that is equivalent to random chance for binary classification. When looking at the fully convolutional model, data augmentation also reduced the standard deviation of accuracy results which indicate that there was a reduction in noise when classifying in the experimental domain and indicates a greater confidence in the statistical outcomes in the experimental domain.

*Table 2: Classification results for the reduction architecture.*

| Model | With/Without Augmentation | Evaluation Metric | | | | | | | |
|---|---|---|---|---|---|---|---|---|---|
| | | Accuracy | | F1 | | Precision | | Recall | |
| | | Mean | Standard Deviation | Mean | Standard Deviation | Mean | Standard Deviation | Mean | Standard Deviation |
| Fully Conv | With | **0.918** | 0.114 | **0.892** | 0.164 | **0.985** | 0.044 | **0.852** | 0.228 |
| | Without | 0.836 | 0.139 | 0.833 | 0.116 | 0.921 | 0.159 | 0.804 | 0.172 |
| Max Pool | With | 0.672 | 0.162 | 0.438 | 0.327 | 0.885 | 0.298 | 0.360 | 0.345 |
| | Without | 0.510 | 0.200 | 0.236 | 0.290 | 0.35 | 0.391 | 0.320 | 0.447 |

### 3.3. Architecture 2 "Constant": Combined reduction and feature extraction

The second architecture, herein known as constant, utilized cuboidal kernels and max pooling or strided convolutional layer with non-uniform dimensions (Figure 15), allowing for feature extraction and dimensionality reduction to be performed consistently throughout the first blocks layers of the network instead of as two distinct processes in the initial architecture. After four blocks, the dimensionality of the data was the same as in the reduction model and as with the first architecture, a feature block with cube kernels of equal dimensionality was used. As with the reduction architecture, global average pooling was used to reduce the number of features for classification, whilst batch normalization and dropout were used in accordance with Figure 14. The final architecture is given by the diagram in Figure 14. The total parameter size of the network was estimated to be 1.23 M parameters when convolutional layers were used for dimensionality reduction, whereas the parameter size reduced to 0.69 M when max pooling was employed.

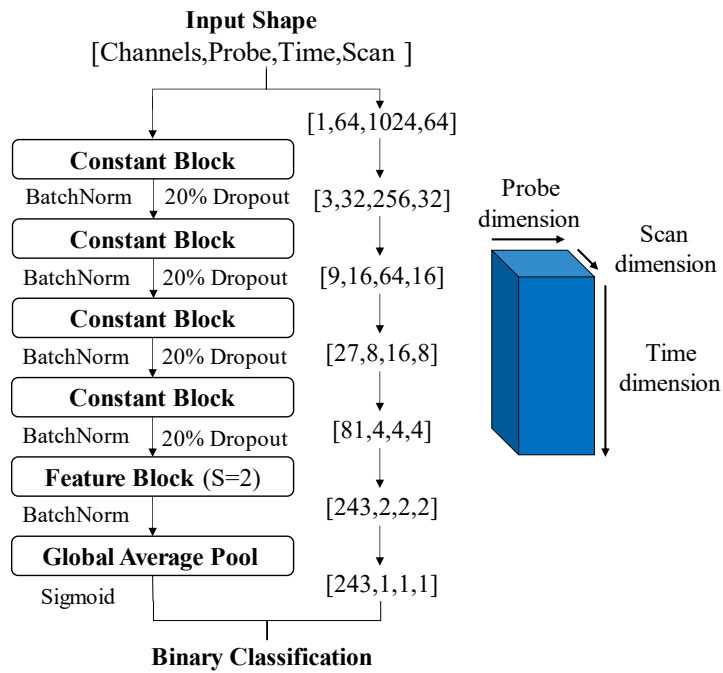

*Figure 14: Network architecture for the constant model.*

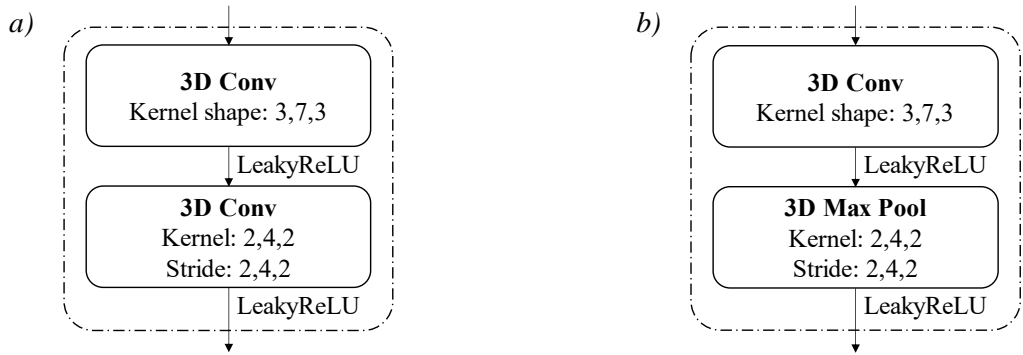

*Figure 15: Constant blocks with either convolution (a) or max pooling (b) layers for dimensionality reduction.*

### 3.3.1. Classification results

Table 2Table 3 provides a summary of classification results for the constant architecture. As with the reduction architecture, whilst the fully convolutional architecture did come at a higher computational cost, it greatly outperformed the max pooling architecture when compared directly with or without data augmentation. It marginally improved on the best performing reduction architecture when augmentation was used and improved on the constant architecture with max pooling by an increase in mean accuracy of 9.0%.

Max pooling proved to be much more robust with the constant architecture compared to the reduction architecture. When trained with and without data augmentation max pooling resulted in models that performed significantly better than random chance which is a clear improvement over the reduction architecture. Furthermore, when using max pooling and data augmentation, the mean classification accuracy was higher than the fully convolutional models without using augmentation for both architecture 1 and 2. As seen in the reduction architecture, the use of data augmentation was highly effective and resulted in over a 10% improvement in accuracy for both models. Augmentation also reduces the standard deviation of accuracy results, again improving confidence in the classification performance in the experimental domain.

*Table 3: Classification results for the constant architecture.*

| Model | With/Without Augmentation | Evaluation Metric | | | | | | | |
|---|---|---|---|---|---|---|---|---|---|
| | | Accuracy | | F1 | | Precision | | Recall | |
| | | Mean | Standard Deviation | Mean | Standard Deviation | Mean | Standard Deviation | Mean | Standard Deviation |
| Fully Conv | With | **0.922** | 0.950 | **0.904** | 0.130 | **0.975** | 0.050 | 0.872 | 0.194 |
| | Without | 0.788 | 0.200 | 0.845 | 0.128 | 0.767 | 0.201 | **0.984** | 0.032 |
| Max Pool | With | 0.846 | 0.143 | 0.785 | 0.230 | 0.853 | 0.300 | 0.748 | 0.290 |
| | Without | 0.712 | 0.198 | 0.490 | 0.420 | 0.574 | 0.473 | 0.444 | 0.404 |

## 3.4. Architecture 3 "NAS": 3D ResNet based Neural Architecture Search
### 3.4.1. Neural Architecture Search

The final architecture was developed through NAS of a modified ResNet search space to account for 3D convolutions and operations. One of the challenges in applying NAS to a new domain task is the design of the search space. For this task, a new search-space framework which utilizes a novel search space based on a ResNet-like structure is introduced. A fixed stem was used to down sample the data by a factor of 4 in the spatial dimensions and a factor of 8 in the time dimension whilst aiming to retain information through increasing the channels to 64. A further down sample block with average pooling followed by two to four residual blocks were all searched individually. An overview of the structure can be seen in Figure 16. The residual blocks and bottleneck features of the ResNet architecture are retained, whilst searching operations for each edge within the residual block. This provided a large diversity of architectures, which is key to attaining good performance in a novel application, whilst also ensuring that many networks conformed to successful design principles. Each residual block contained two fixed point-wise convolutions used to down and up sample the number of channels. Figure 17 shows an example of a residual block denoting the searched and fixed operations.

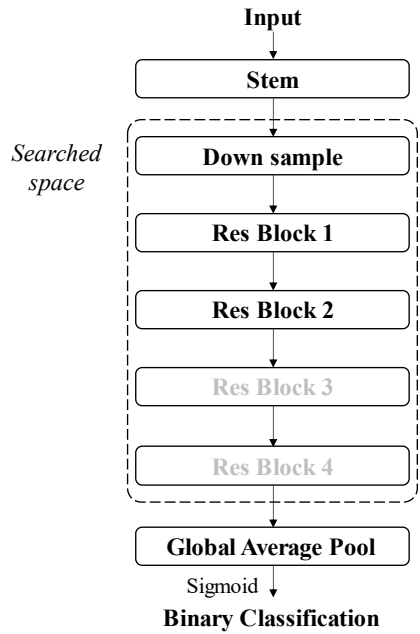

*Figure 16: Representation of the ResNet style searched space.*

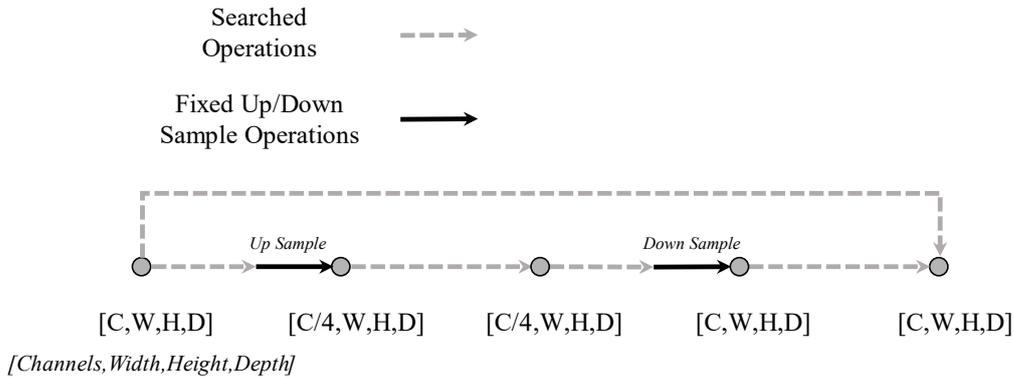

*Figure 17: Diagram of the searched residual block.*

These blocks were then stacked in groups, with the resolution down sampled between groups. Equation 5 gives the probability of a new group being created for each residual block, otherwise they were added to the current group. This makes groups unlikely to be extremely long or short.

$$P(newResBlockGroup) = \frac{1}{currentResBlockGroupSize} \qquad 5$$

The primitive operations of a search space are the list of operations which are assigned to the edges of a network architecture. The implemented approach incorporated a standard set of operations commonly found in the NAS literature. These operations comprised of convolutions, pooling, and skip connections, which are widely recognized and utilized within the field. These operations were all 3D due to the dimensionality of the data. In contrast to standard practice, which makes use of separable convolutions, the approach presented in this study deployed both depth-wise and point-wise convolutions as the fundamental convolutions within the search space. This significantly reduced the number of parameters in each operation of the architecture, greatly reducing the computational cost. Specifically, the depth-wise convolutions were applied with equidimensional cube kernels, of size 3, 5, or 7, coupled with dilation values that ranged from 1 to 4. Skip connections, point-wise convolutions, as well as average and max pooling operations were also searched for. For the pooling operations, equidimensional cube kernels of size 3, 5, or 7, with a dilation value of one were employed. The search encompassed the exploration of GELU activation function and batch normalization, as well as the absence of activation and normalization operations. This allowed for architectures with fewer activation and normalization function which has been shown to be beneficial [50]. The searched down sample operation had a fixed kernel size with two in the spatial dimensions a four in the temporal dimension with a dilation of one. Throughout the relevant operations, a stride of one was employed.

A simple random search was applied to this search space for 80 iterations. Each model was evaluated using the validation dataset every 10 batches, with the lowest loss on validation across the training taken as the evaluation metric. For each searched architecture, a model was retrained with new initializations three times and the mean evaluation metrics were used when selecting the discovered architecture, this ensured a more accurate estimate of model performance. Cross validation was unable to be used as the combination of NAS and domain transfer would have resulted in data leak between the NAS stage and the final model test evaluation stage. Figure 18 provides an overview of the NAS process and demonstrates how separation of the validation and test set were maintained in context of the complete model pipeline, given in Figure 3.

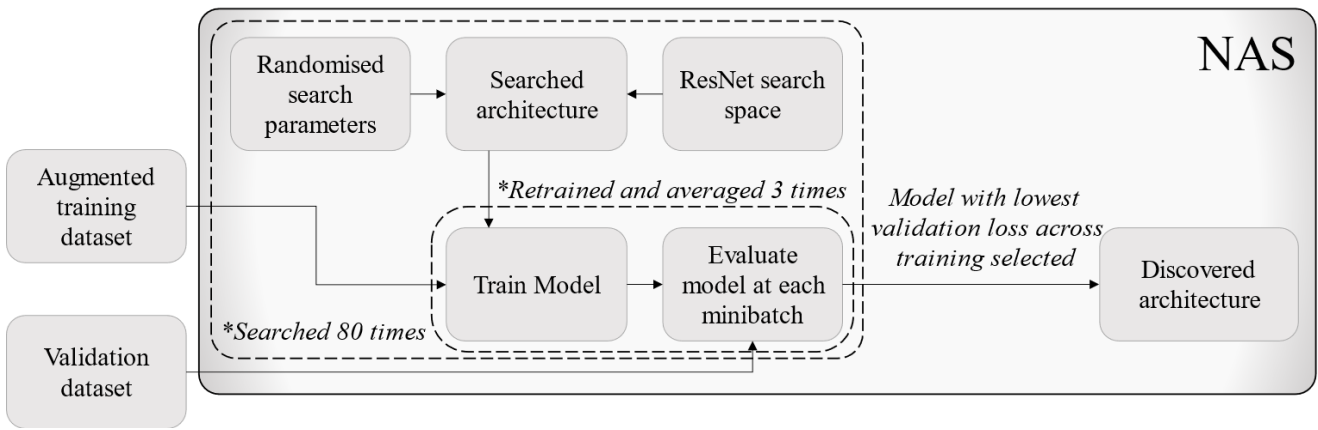

*Figure 18: Overview of the process for NAS implementation*

The final discovered architecture had 1.03 M parameters and is given in Figure 19, with the details of the residual blocks given in Figure 20.

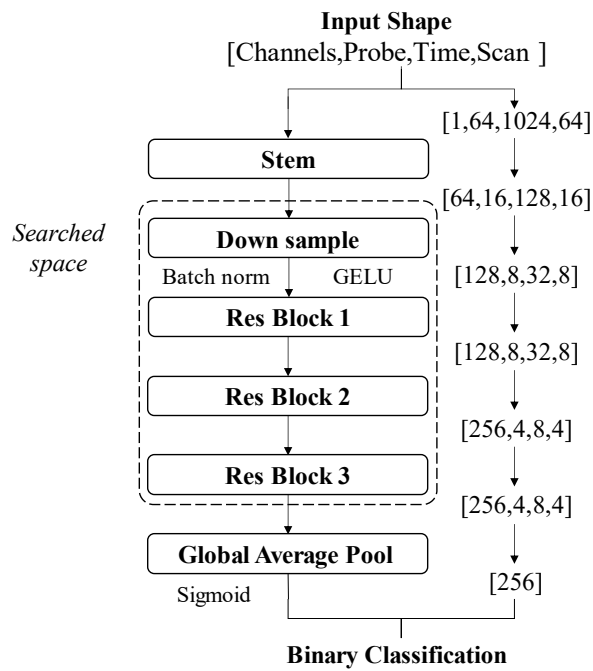

*Figure 19: The overall structure of the discovered architecture.*

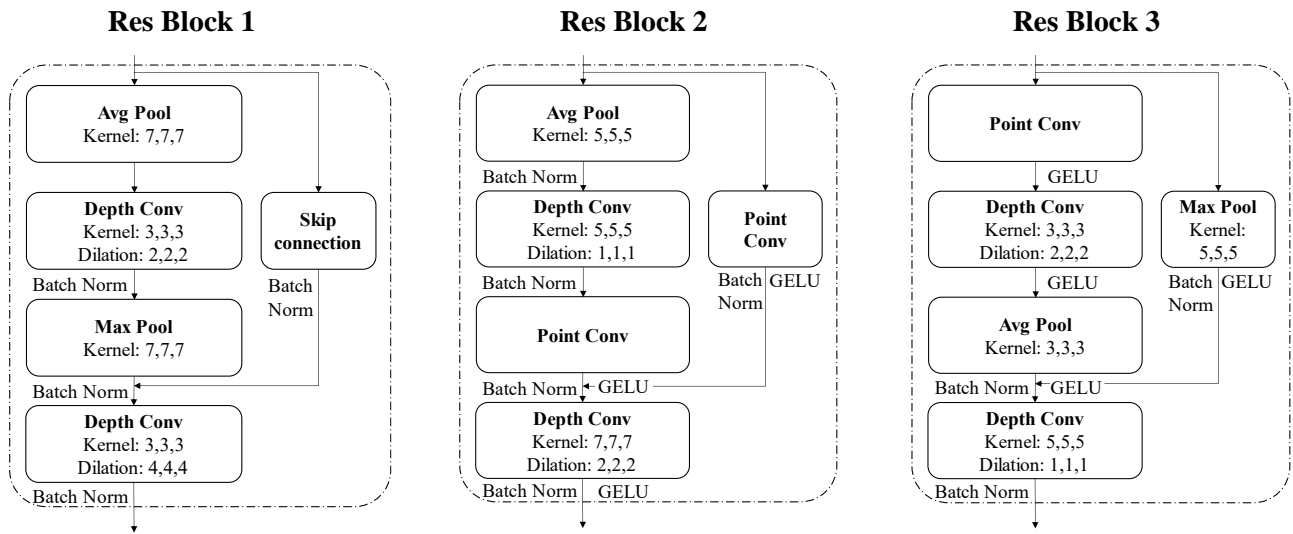

*Figure 20: The details of each discovered residual block.*

### 3.4.2. Classification results

Table 4 summarizes the classification results for the architecture discovered by NAS. The architecture consistently produced ideal results when trained using data augmentation, with a mean classification accuracy of 1.00, with 0.00 standard deviation across the 10 separate training iterations, this demonstrated high confidence in the model's conclusions in the target domain. Discarding data augmentation during training had a significant impact on the classification performance with a 22.4% drop in mean accuracy, along with a standard deviation increase in 17.8%, which demonstrated a significant reduction in statistical confidence.

*Table 4: Classification results for the searched architecture*

| Model | With/Without Augmentation | Evaluation Metric | | | | | | | |
|---|---|---|---|---|---|---|---|---|---|
| | | **Accuracy** | | **F1** | | **Precision** | | **Recall** | |
| | | Mean | Standard Deviation | Mean | Standard Deviation | Mean | Standard Deviation | Mean | Standard Deviation |
| NAS | With | **1.00** | 0.00 | **1.00** | 0.00 | **1.00** | 0.00 | **1.00** | 0.00 |
| | Without | 0.776 | 0.178 | 0.830 | 0.123 | 0.745 | 0.190 | 0.972 | 0.044 |

## 4. Discussion

Figure 21 shows a comparison of classification performance, with mean and standard deviation of accuracies for each model, including the effects of augmentation during training. Both the reduction and constant designed architectures were able to achieve similar max classification accuracies when using fully convolutional architectures and data augmentation with mean results of 91.8% and 92.2% respectively. For this instance, the constant architecture showed a lower standard deviation of accuracies by 1.9% which indicated a slight increase in statistical confidence compared to the reduction model. The constant model was far more robust when using max pooling for dimensionality reduction compared to the reduction model. The significance of this is demonstrated by the fact that the reduction model produced a poorer classifier using max pooling with augmentation than the constant model without augmentation. This indicates that using cuboidal kernels of non-equal lengths was effective when combining dimensionality reduction and feature extraction for extreme aspect ratios.

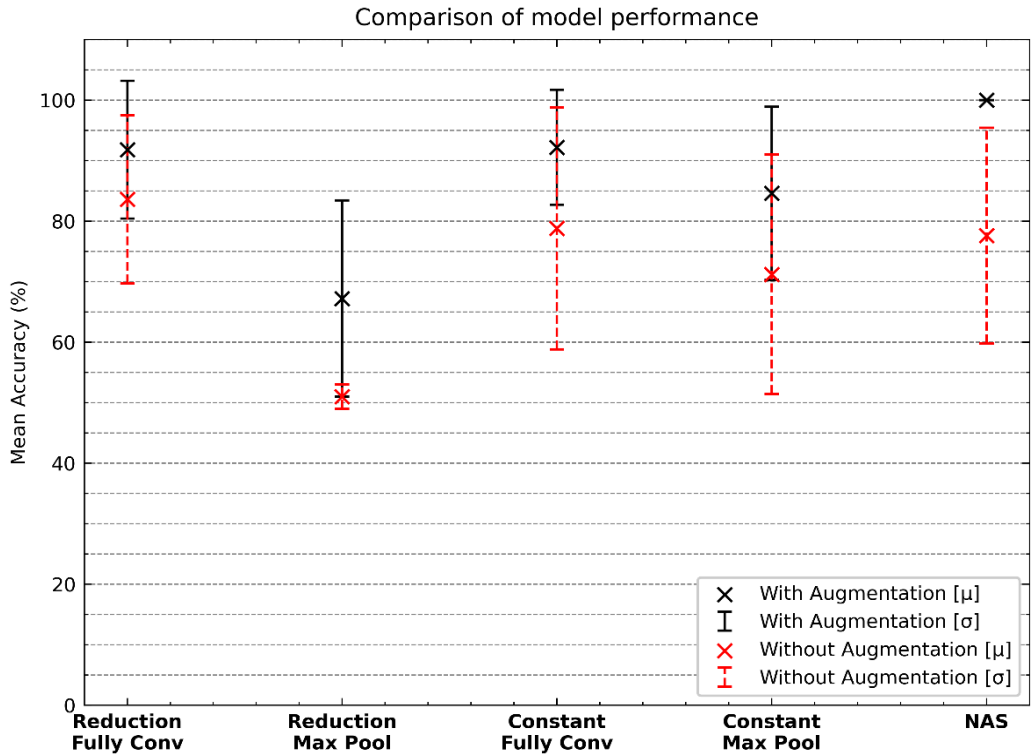

*Figure 21: Comparison of classification results for each model.*

The experimental results demonstrated that the architecture generated from NAS greatly outperformed the other two in terms of classification accuracy when using data augmentation. Whilst all the models used in this paper are not large and are considerably smaller than typical sizes for 2D ResNet's and other CNNs [37], the NAS model was able to achieve the highest performance with a lower computational cost, at only 67% of the total parameter size of the best performing reduction model and 80% of the total parameter size of the best performing constant model. This demonstrated the importance of utilizing neural architecture search to optimize CNNs. However, when trained without data augmentation the NAS model performed significantly worse and was outperformed by both fully convolutional models with and without augmentation. This indicates that this model's architecture relied even more heavily on effective data augmentation.

Whilst ideal classification was achieved consistently for the discovered architecture when trained with data augmentation, this was tested on detection of manufactured defects only. Specifically, back drilled holes which are perpendicular to the propagating sound wave and act as ideal reflectors. This makes them comparably easier to detect than other defects. Whilst samples with naturally occurring defects are challenging to get access to, in future work the authors aim to expand the simulation scope and test the models on naturally occurring defects which will likely prove more challenging to detect. For more challenging detection and characterization tasks a more sophisticated search optimization algorithm could be employed to discover architectures more efficiently.

The achieved classification results suggest that the synthetic data generation process is a viable approach for producing fully synthetic 3D UT volumetric datasets that closely map to the experimental domain and enable the development of effective classifiers. However, due to the substantial improvement in classification performance achieved through the implementation of data augmentation methods, it is important to acknowledge that disparities between the synthetic and experimental domains persist. This observation underscores the necessity for augmentation techniques to further enhance the generalizability of the model. Nonetheless, it is worth noting that the data augmentation methods employed in this study proved to be highly effective in aiding not only generalizability but also in facilitating the transfer of knowledge across domains.

The key benefits for analyzing the complete 3-D volumetric data instead of processed images were the ability to learn from greater features, the reduction in pre-processing requirements, and the potential reduction in inference time by analyzing the complete volume all at once. The impact on inference time is challenging to quantify, however if comparing the compute required to process 64 B-scan images (the equivalent spatial scan data), without parallelization for equivalent 2D classifiers, there is the potential for up to 64 times saving in inference time for the same scan area. Despite these advantages there are still potential benefits to analyzing UT data as images. One of these is the many opportunities for detection of a single defect in multiple B-scans. It is likely that defects will span multiple B-scan images, and as such by analyzing each B-scan there are multiple chances to detect an individual defect. This means an individual defect can still be detected even if individual defective images are incorrectly classified. However, the opportunities for characterization and localization of defects are far greater when retaining the volumetric spatial information and this work opens future prospects for 3D classification and segmentation which would be much more challenging if using C-scans or B-scans alone.

The research outcomes demonstrated the considerable potential of employing 3D-CNNs in conjunction with well-designed data augmentation techniques and optimized architecture search spaces to address challenging 3D classification tasks characterized by extreme aspect ratios, as observed in the context of UT. Insufficient utilization of data augmentation severely hampered the model's ability to generalize to experimental datasets, leading to suboptimal classification performance. Likewise, choosing an unsuitable model architecture could result in the failure to capture crucial features necessary for accurate classification. Consequently, it is imperative to thoroughly consider both aspects during the design of a classification model for 3D UT data to ensure optimal performance.

## 5. Conclusion

Deep learning has demonstrated prior success in ultrasonic non-destructive evaluation when applied to either time series or image data. However, analyzing only time series or image data can result in a significant loss of information in either the temporal or spatial domains. This paper proposes the use of 3D convolutional neural networks to classify complete volumetric ultrasound data without compression, retaining all spatial and temporal information. This approach not only reduced the need for accurate gating when constructing C-scan images but also decreased the amount of signal processing required. To train the models, synthetic data was generated from semi-analytical simulations, while experimentally collected ultrasonic responses from manufactured defects were used for testing. Two forms of data augmentation were implemented based on physical variations seen in experimental ultrasonic responses to improve the model's classification performance in the experimental domain. Furthermore, the performance of three different architectures, including two hand designed architectures and one designed by Neural Architecture Search (NAS) from a ResNet search space modified for 3D, were compared.

The constant model achieved similar results to the reduction model in the best case with mean classification accuracies of 92.2% and 91.8% respectively. However, the constant model resulted in a more robust architecture which still demonstrated limited effectiveness when using max pooling instead of a fully convolutional model. The third architecture, designed by NAS, when trained with data augmentation, gave the best results, providing 100% classification accuracy. However, this performance was highly dependent on data augmentation during training and would perform worse than the other fully connected models if data augmentation was not used. Overall, data augmentation had a significant impact on model performance and always resulted in a large increase in model performance, ranging from an 8.2% to 22.4% increase in mean accuracy.

Overall, this work demonstrated that it is possible to train successful Deep Learning models to classify full volumetric ultrasonic data for NDE. The issue of a lack of data in most NDE situations was addressed by successfully implementing synthetic data generation in 3D. The work highlighted the importance of appropriate architecture selection and effective data augmentation when translating between synthetic and experimental domains, with both factors essential in achieving high classification accuracy.

The focus of this work was on the use of volumetric datasets, and whilst 100% classification accuracy was achieved through effective NAS, the authors recognize that back drilled holes are generally simple defects to detect by human operators. Future work aims to increase the complexity of the task by detecting a wider range of more challenging defects and expanding the simulation scope to better cover naturally occurring defects.

# 6. Acknowledgment

This work was supported through Spirit AeroSystems/ Royal Academy of Engineering Research Chair for In-Process Non-Destructive Testing of Composites, RCSRF 1920/10/32.